\documentclass[aps,twocolumn,superscriptaddress,showpacs]{revtex4}

\usepackage[dvips]{graphicx}
\newcommand{\nc}{\newcommand}           
\nc{\vc}[1]     {\mbox{\boldmath $#1$}} 
\nc{\bs}[1]     {\boldsymbol{#1}}
\nc{\wtil}      {\widetilde}            
\nc{\bra}       {\langle}               
\nc{\ket}       {\rangle}               
\nc{\bras}[1]   {\langle #1|}           
\nc{\kets}[1]   {|#1\rangle}            
\nc{\hO}        {\hat{O}}               
\nc{\mapleft}[1]{                       
 \smash{\mathop{                      %
  \hbox to 0.90cm{\rightarrowfill} }\limits_{#1}}}

\def\JL#1#2#3#4{ {{\rm #1}} \textbf{#2}, #4 (#3).}  
\nc{\PR}[3]     {\JL{Phys. Rev.}{#1}{#2}{#3}}
\nc{\PRC}[3]    {\JL{Phys. Rev.~C}{#1}{#2}{#3}}
\nc{\PRA}[3]    {\JL{Phys. Rev.~A}{#1}{#2}{#3}}
\nc{\PRL}[3]    {\JL{Phys. Rev. Lett.}{#1}{#2}{#3}}
\nc{\NP}[3]     {\JL{Nucl. Phys.}{#1}{#2}{#3}}
\nc{\NPA}[3]    {\JL{Nucl. Phys.}{A#1}{#2}{#3}}
\nc{\PL}[3]     {\JL{Phys. Lett.}{#1}{#2}{#3}}
\nc{\PLB}[3]    {\JL{Phys. Lett.~B}{#1}{#2}{#3}}
\nc{\PTP}[3]    {\JL{Prog. Theor. Phys.}{#1}{#2}{#3}}
\nc{\PTPS}[3]   {\JL{Prog. Theor. Phys. Suppl.}{#1}{#2}{#3}}
\nc{\PRep}[3]   {\JL{Phys. Rep.}{#1}{#2}{#3}}
\nc{\JP}[3]     {\JL{J. of Phys.}{#1}{#2}{#3}}
\nc{\andvol}[3] {{\it ibid.}\JL{}{#1}{#2}{#3}}
\nc{\mydraft}	{\setlength{\topmargin}{-1.5cm}}
\mydraft

\begin{document}

\title{One-neutron removal strength of $^7$He into $^6$He using the complex scaling method
}

\author{Takayuki Myo\footnote{myo@ge.oit.ac.jp}
}
\affiliation{
General Education, Faculty of Engineering, Osaka Institute of Technology, Osaka 535-8585, Japan}
\affiliation{
Research Center for Nuclear Physics (RCNP), Osaka University, Ibaraki 567-0047, Japan}

\author{Ryosuke Ando\footnote{ando@nucl.sci.hokudai.ac.jp}
}
\affiliation{
Division of Physics, Graduate School of Science,
Hokkaido University, Sapporo 060-0810, Japan}

\author{Kiyoshi Kat\=o\footnote{kato@nucl.sci.hokudai.ac.jp}
}
\affiliation{
Division of Physics, Graduate School of Science,
Hokkaido University, Sapporo 060-0810, Japan}

\date{\today}

\begin{abstract}
We study the one-neutron removal strength of the $^7$He ground state, which provides us with the $^6$He-$n$ component in $^7$He.
The He isotopes are described on the basis of the $^4$He+$Xn$ cluster model ($X=1,2,3$).
The complex scaling method is applied to describe not only the Gamow resonances 
but also the nonresonant continuum states of valence neutrons, with the correct boundary condition of particle decays.
The one-neutron removal strength of $^7$He into the unbound states of $^6$He is calculated
using the complex-scaled Green's function, in which a complex-scaled complete set of $^4$He+$n$+$n$ states is adopted.
Using this framework, we investigate resonant and nonresonant contributions of the strength,
which individually produce specific structures in the distributions.
In addition, we propose a method to obtain the real-value strength using the complex values of spectroscopic factors of Gamow states.
As a result, the $^6$He($2^+$) resonance is found to give the largest contribution.
\end{abstract}

\pacs{
21.60.Gx,~
21.10.Pc,~
27.20.+n~
}

\maketitle 

\section{Introduction}

The development of experiments using radioactive beam has provided us 
with much information on unstable nuclei far from the stability.
In particular, the light nuclei near the drip-line exhibit new phenomena of nuclear structures,
such as the neutron halo structure in $^6$He and $^{11}$Li \cite{Ta92,Ta96}.

Recently, many experiments \cite{Ko99,Bo01,Me02,Wu05,Bo05,Sk06,Ry06,Be07,Wu08} on $^7$He have been reported
and have confirmed that its ground state is assigned to be the $3/2^-$ resonant state 
at 0.3$\sim$0.5 MeV above the $^6$He+$n$ threshold energy.
For the excited states, $^7$He can decay not only to two-body $^6$He+$n$ channels, 
but also to many-body channels of $^5$He+2$n$ and $^4$He+3$n$, 
because $^6$He is a Borromean nucleus and breaks up easily into $^4$He+$n$+$n$. 
This multiparticle decay condition makes it difficult to settle the excited states of $^7$He.
In fact, contradictions in the observed energy levels and their spins still remain.
The excited state at $E_x \sim 3$ MeV has been assigned as $5/2^-$ in several experiments \cite{Ko99,Bo01,Wu05,Sk06,Ry06}.
The existence of $1/2^-$ is also expected \cite{Me02,Bo05,Wu05,Sk06,Ry06}, but the energies and decay width are not yet fixed.
This $1/2^-$ state with a possibility of the $LS$ partner of the ground state is of interest
because the $LS$ splitting energy in $^7$He gives information on the $LS$ interaction in drip-line nuclei.
Some experiments \cite{Me02,Sk06,Ry06} report $1/2^-$ at around 1 MeV in the low excitation energy. 
However, other observations \cite{Wu05,Bo05} suggest a little higher excitation energy.
The spectroscopic factor associated with the $^6$He halo state has also been reported \cite{Be07}.

On the theoretical side, {\it ab initio} calculations of the no-core shell model \cite{Na98} and the Green's function Monte Calro \cite{Pi04}
have been performed, and the calculated energies of the ground and $5/2^-$ states show a good correspondence with the experiments.
The $1/2^-$ state is predicted at around 3 MeV, although the results depend on the choice of the three-nucleon forces \cite{Pi04}. 
Those calculations are based on the bound state approximation and the continuum effect from many-body open channels is not included, 
while the observed $^7$He states are unbound.

Several methods have been proposed to treat the continuum effects explicitly, 
such as the continuum shell model \cite{Vo05} and the Gamow shell model \cite{Ha05,Mi07,Ha07}.
It is, however, difficult to satisfy the multiparticle decay conditions correctly for all open channels. 
The obtained energy spectra and decay widths of He isotopes depend on the treatment of open channels.
For the spectroscopy of $^7$He, it is necessary to describe the four-body resonances of the $^4$He+3$n$ system in the theoretical model.
Furthermore, it is important to reproduce the threshold energies of the particle decays into subsystems.
Emphasizing these theoretical conditions, in our previous work \cite{My07c}, 
we investigated the $^7$He spectroscopy with the appropriate treatment of the decay properties concerned with the subsystems of $^{5,6}$He.
We employed the cluster orbital shell model (COSM) of $^4$He+$Xn$ \cite{Su88,Ma06,Ma07}, 
in which the open channel effects of the $^6$He+$n$, $^5$He+2$n$ and $^4$He+3$n$ decays are taken into account explicitly.
This means that our analysis satisfies the simultaneous descriptions of $^{4,5,6}$He.
We described the many-body resonances by using the complex scaling method (CSM) \cite{ABC,Ho83,Mo98,Ao06},
under the correct boundary conditions for all decay channels. 
In CSM, the energies and decay widths of many-body resonances are directly obtained by 
diagonalization of the complex-scaled Hamiltonian with $L^2$ basis functions.

We found five resonances of $^7$He and investigated their properties,
such as energies, decay widths, configurations and spectroscopic factors ($S$ factors) of the $^6$He-$n$ components.
In particular, $S$ factors are important in understanding the coupling between $^6$He and a last neutron in $^7$He.
In the analysis of the $S$ factors of the Gamow states of $^{6,7}$He, the $S$ factors are often described in complex values,
because Gamow states have complex eigenvalues and their matrix elements are also complex values \cite{Ao06,Be96}.
These complex $S$ factors lead to the problem of physical interpretation with respect to the observation, 
which was discussed in a previous study \cite{My07c}.
To avoid this problem, in this article, we develop the method of considering 
the contributions of nonresonant continuum states for $S$ factors in addition to those of Gamow states.
This is performed by employing the complete set consisting of Gamow resonant and nonresonant continuum states 
in the continuum energy region.
Using this complete set of unbound states of $^6$He, 
we show the one-neutron removal strength of the $^7$He ground state into the $^6$He unbound states.
This strength function reflects the $^6$He-$n$ components of $^7$He and is observable.

\begin{figure}[t]
\centering
\includegraphics[width=7.5cm,clip]{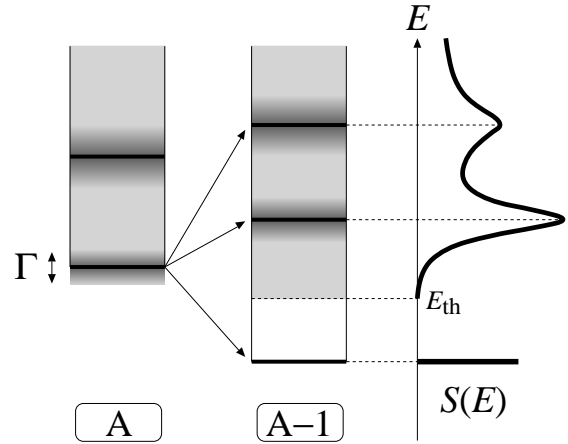}
\caption{Schematic illustration of the one-nucleon removal strength $S(E)$, 
from the mass $A$ system to the $A-1$ one. The excitation energy $E$ is for the $A-1$ system.
$E_{\rm th}$ is the lowest threshold energy of particle emission for the $A-1$ system.}
\label{fig:sche}
\end{figure}

We explain the relation between the one-neutron removal strength and the $S$ factors.
In Fig.~\ref{fig:sche}, we depict the schematic illustration of the one-nucleon removal strength $S(E)$
from a mass $A$ system to an $A-1$ one.
The strength function $S(E)$ involves the information of bound, resonant, and nonresonant continuum states 
of the $A-1$ final states.
The resonant parts of the strength correspond to the $S$ factors of resonances.
Beyond the threshold energy $E_{\rm th}$, the unbound states start to appear, and the strength becomes continuous.
It is, however, difficult to describe the unbound states beyond the two-body case, like $^4$He+$n$+$n$ of $^6$He.
In the previous study \cite{My07c}, only the resonant component ($S$ factors) of the final states is considered 
as a part of $S(E)$, and the nonresonant component is missing.
Because of this partial treatment of the unbound states, 
the relation between the complex $S$ factors and the real observable $S(E)$ is obscure.
Hence, in this study, we include the contributions not only of the resonances, 
but also of the nonresonant continuum states in the distribution, $S(E)$ in Fig.~\ref{fig:sche}.
Some resonances of the $A-1$ system can make structures in $S(E)$ like those shown in Fig.~\ref{fig:sche}.
It is interesting to clarify the origins of the structures seen in the strength.

In this article, we calculate the one-neutron removal strength of $^7$He 
by using the complex-scaled $^4$He+$n$+$n$ complete set of the $^6$He final states \cite{My01}, 
which consists of not only the $^6$He resonances, but also the continuum states of $^5$He+$n$ and $^4$He+$n$+$n$.
We express the $^6$He complete set by using the complex-scaled solutions of the $^4$He+$n$+$n$ model.
It has been shown that CSM is a powerful method to investigate the resonances and the nonresonant continuum states.
So far, using CSM, the Coulomb breakup strengths of halo nuclei have been successfully analyzed \cite{Ao06,My01,Su02,My03,My07b,My08}.
In calculations of the strength $S(E)$ in this study, 
we adopt the complex-scaled Green's function described by using the complex-scaled eigenstates of $^6$He.
Similar methods using complex-scaled Green's function have been used successfully in solving the nuclear reaction problems \cite{Kr07,Su08,Ki09}.

In this study, we evaluate the one-neutron removal strength of $^7$He into $^6$He as an observable.
This means that we develop a method to obtain the observable from the complex $S$ factors of Gamow states.
Furthermore, we investigate the structures in the strength
by decomposing the $^6$He unbound states into $^6$He resonances, $^5$He+$n$ and $^4$He+$n$+$n$ continuum components \cite{My01}.
As was shown in the Coulomb breakups of halo nuclei, 
the contributions of each component can be unambiguously classified by using CSM.
This is a prominent point of the present method.

In Sec.~\ref{sec:model}, we explain the complex-scaled COSM wave function for He isotopes, 
and the method of calculating the neutron removal strength using CSM by introducing the complex-scaled Green's function.
In Sec.~\ref{sec:result}, we discuss the $^7$He structures, the $S$ factors of the $^6$He-$n$ components, and 
also the one-neutron removal strength into $^6$He.
A summary is given in Sec.~\ref{sec:summary}.
\section{Complex-scaled $^4$He+\vc{XN} COSM for He isotopes}
\label{sec:model}

\subsection{COSM for the $^4$He+\vc{Xn} systems}

We explain the COSM of the $^4$He+$Xn$ systems, in which $X=1$ for $^5$He, $X=2$ for $^6$He and $X=3$ for $^7$He.
The Hamiltonian is the same as that used in Refs.~\cite{My07c,My01};
\begin{eqnarray}
	H
&=&	\sum_{i=1}^{X+1}{t_i} - T_G +	\sum_{i=1}^{X}V^{\alpha n}_i + \sum_{i<j}^X V^{nn}_{ij},
        \label{eq:Ham}
\end{eqnarray}
where $t_i$ and $T_G$ are the kinetic energies of each particle ($Xn$ and $^4$He) and of the center of mass of the total system, respectively.
The interactions $V^{\alpha n}$ and $V^{nn}$ are given by the modified KKNN potential \cite{Ao95} and the Minnesota potential \cite{Ta78}, respectively. 
They reproduce the low-energy scattering data of the $^4$He-$n$ and the $n$-$n$ systems, respectively.

\begin{figure}[t]
\centering
\includegraphics[width=8.0cm,clip]{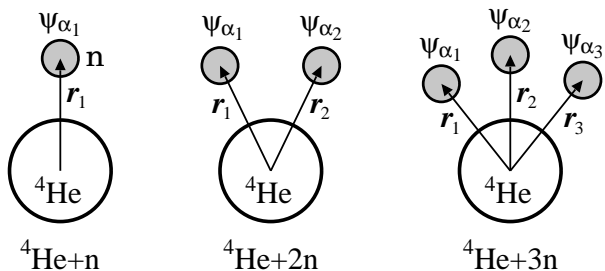}
\caption{Sets of the spatial coordinates in the COSM for the $^4$He+$Xn$ system.}
\label{fig:COSM}
\end{figure}
 
For the wave function, $^4$He is treated as the $(0s)^4$ configuration of a harmonic oscillator wave function, 
whose length parameter $b_c$ is 1.4 fm to fit the charge radius of $^4$He.
The motion of valence neutrons around $^4$He is solved variationally using the few-body technique.
We expand the relative wave functions of the $^4$He+$Xn$ system using the COSM basis states \cite{Su88,Ma06,Ma07}.
In COSM, the total wave function $\Psi$ of the $^4$He+$Xn$ system is given by the superposition of the configuration $\Psi_c$ as
\begin{eqnarray}
    \Psi(^{4}{\rm He}+Xn)
&=& \sum_c C_c \Psi_c(^{4}{\rm He}+Xn),
    \label{WF0}
    \\
    \Psi_c(^{4}{\rm He}+Xn)
&=& \prod_{i=1}^X a^\dagger_{\alpha_i}|0\rangle, 
    \label{WF1}
\end{eqnarray}
where $^4$He corresponds to a vacuum $|0\rangle$.
The creation operator $a^\dagger_{\alpha_i}$ is for the valence neutron above $^4$He,
with the quantum number $\alpha_i$ in a $jj$ coupling scheme. 
The index $i=1,\cdots,X$ is for $X$ valence neutrons. The set of $\alpha_i$ is included in the index $c$.
The variational coefficient is expressed by $C_c$ with respect to $\Psi_c$.
We take a summation over the available configurations.
The coordinate representation of the single-particle state corresponding to $a^\dagger_{\alpha_i}$ is given as 
$\psi_{\alpha_i}$ with the relative coordinate $\vc{r}_i$ between $^4$He and a valence neutron shown in Fig.~\ref{fig:COSM}.
Considering the angular momentum coupling, the total wave function $\Psi^J$ with the spin $J$ is expressed as
\begin{eqnarray}
    \Psi^J(^{4}{\rm He}+Xn)
&=& \sum_c C_c^J \Psi^J_c(^{4}{\rm He}+Xn),
    \label{WF3}
    \\
    \Psi^J_c(^{4}{\rm He}+Xn)
&=& {\cal A}^\prime \left\{\, [\Phi(^4{\rm He}), \chi^{J}_c(Xn)]^J\, \right\},
    \label{eq:WF}
    \\
    \chi^{J}_c(n)
&=& \psi_{\alpha_1}^J,
    \\
    \chi^{J}_c(2n)
&=& {\cal A}\{ [\psi_{\alpha_1},\psi_{\alpha_2}]_J \},
    \label{eq:WF6}
    \\
    \chi^{J}_c(3n)
&=& {\cal A}\{ [[\psi_{\alpha_1},\psi_{\alpha_2}]_{j_{12}},\psi_{\alpha_3}]_J \}.
    \label{eq:WF7}
\end{eqnarray}
Here, as shown in Fig. \ref{fig:COSM}, $\chi^J_c(Xn)$ expresses the wave functions for valence neutrons.
The spin $j_{12}$ is a coupled angular momentum of the first and second valence neutrons. 
The antisymmetrizers between valence neutrons and
between a valence neutron and neutrons in $^4$He are expressed as ${\cal A}$ and ${\cal A}^\prime$, respectively.
The effect of ${\cal A}^\prime$ is treated in the orthogonality condition model \cite{Ao06,My01,Ma07}, in which
$\psi_{\alpha}$ is imposed to be orthogonal to the $0s$ state occupied by $^4$He.
The radial part of $\psi_\alpha$ is expanded with a finite number of Gaussian basis functions as
\begin{eqnarray}
    \psi_\alpha
&=& \sum_{k=1}^{N_\alpha} C_{\alpha,k}\ \phi_{\alpha}^k(\vc{r},b_{\alpha,k}),
    \label{WFR}
    \\
    \phi_{\alpha}^k(\vc{r},b_{\alpha,k})
&=& {\cal N} r^{\ell_\alpha} e^{-(r/b_{\alpha,k})^2/2} [Y_{\ell_\alpha}(\hat{\vc{r}}),\chi^\sigma_{1/2}]_{j_\alpha}.
\end{eqnarray}
Here, the index $k$ is for the Gaussian basis with the length parameter $b_{\alpha,k}$.
A basis number for the state $\alpha$ and the normalization factor for the basis
are given by $N_\alpha$ and ${\cal N}$, respectively. 
Two expansion coefficients $\{C_c^J\}$ and $\{C_{\alpha,k}\}$ are determined variationally 
with respect to the total wave function $\Psi^J$ by diagonalization of the Hamiltonian matrix elements.
The length parameters $b_{\alpha,k}$ are chosen as geometric progression \cite{Ao06}.
Due to this expansion of the radial components using a finite number of basis states, 
all the energy eigenvalues are discretized for bound, resonant and continuum states.
We use at most 17 Gaussian basis functions with a maximum range of 40 fm.

Here, we discuss the coupling between $^4$He and valence neutrons.
This is related to the boundary condition of the neutron emission,
which is important, in particular, when we deal with the weakly binding, resonant and continuum states \cite{My07c,My08}. 
We consider the coupling between $^{7}$He and the $^6$He+$n$ configurations. 
Asymptotically, when the last neutron is located far away from $^6${He} 
($\vc{r}_3\to\infty$ for $^4$He+$3n$ in Fig.~\ref{fig:COSM}),
any coupling between $^6$He and a last neutron disappears, and $^6$He becomes its isolated eigenstate
of the Hamiltonian in Eq.~(\ref{eq:Ham}).
\begin{eqnarray}
        \Psi^J(^{7}{\rm He})
&=&     \sum_c C_c^J  {\cal A}^\prime\left\{ [ \Phi(^{4}\mbox{He}), \chi_c^J(3n) ]^J \right\}
        \label{asympt0}
        \\
&\mapleft{\vc{r}_3\to\infty}&
        \left[ \Psi^{J^\prime}_\nu(^{6}\mbox{He}), \psi_{\alpha_3} \right]^{J},
        \label{asympt1}
        \\
        \Psi^{J^\prime}_\nu(^{6}\mbox{He})
&=&     \sum_c C_{c,\nu}^{J^\prime} {\cal A}^\prime \left\{ [ \Phi(^{4}\mbox{He}), \chi_{c,\nu}^{J^\prime}(2n) ]^{J^\prime} \right\},
        \label{asympt2}
\end{eqnarray}
where the spin $J$ and $J'$ are for $^7$He and $^6$He, respectively,
and the index $\nu$ indicates the eigenstate of $^6$He.
The mixing coefficient $\{C^{J^\prime}_{c,\nu}\}$ and the two-neutron wave function $\chi^{J^\prime}_{c,\nu}(2n)$ in Eq.~(\ref{asympt2}) 
are those of the $^6$He eigenstates.
Hence, the three-neutron wave function $\chi^{J}_c(3n)$ in Eq.~(\ref{asympt0}) satisfies the following asymptotic forms
\begin{eqnarray}
    \sum_c C_c^J \chi^{J}_c(3n)
&\mapleft{\vc{r}_3\to\infty} & \left( \sum_c C_{c,\nu}^{J^\prime} \chi^{J^\prime}_{c,\nu}(2n)\right) \psi_{\alpha_3} .
    \label{asympt3}
\end{eqnarray}
This relation implies that the three-neutron wave function of $^7$He 
is asymptotically decomposed into two-neutron wave function of $^6$He and a last neutron. 
Equations~(\ref{asympt0})-(\ref{asympt3}) describe the boundary conditions of the COSM wave functions for He isotopes. 
In contrast, when a last neutron comes close to $^6$He, this last neutron dynamically couples to the $^6$He eigenstates $\Psi_\nu^{J^\prime}$.
This coupling depends on the relative distance between $^6$He and a last neutron,
and changes the $^6$He+$n$ configurations from the eigenstates of $^6$He to the eigenstates of $^7$He.
In the COSM, the structure change of $^6$He inside $^{7}$He is determined variationally to optimize the $^{7}$He eigenstate.
This discussion of asymptotic condition can be also applied for the configurations of $^5$He+$2n$ and $^4$He+$3n$. 
Hence, in the COSM, we can treat the neutron emissions with the correct boundary conditions.

For the single-particle states $\alpha=\ell_j$ $(j=\ell\otimes\frac12)$, 
we take angular momenta $\ell\le 2$ to keep the accuracy of the converged energy within 0.3 MeV. 
In the calculation of $^7$He, we adjust the calculated energies of $^6$He by taking the 178.8 MeV of the repulsive strength 
of the Minnesota force \cite{Ta78} and the three-cluster interaction $V^{\alpha n n}$ for the $^4$He-$n$-$n$ system \cite{My01}. 
The former adjustment of the $NN$ interaction can be understood from the pairing correlation between valence neutrons 
with higher angular momenta $\ell>2$ \cite{Ao95}. 
The latter is considered to come from dominantly the tensor correlation in $^4$He.
Recently, we showed that the binding energy and the excited states of $^6$He can be well explained 
without the three-cluster interaction by taking into account the tensor correlation of $^4$He explicitly \cite{My05,My05b}. 
Following previous studies \cite{My07c,My01}, we use the three-cluster potential:
\begin{eqnarray}
	V^{\alpha n n}
=	\sum_{i<j} v_3\ e^{-({\bf r}_i^2+{\bf r}_j^2)/b_c^2}\,\quad \mbox{with}~~
	v_3
=	-25~{\rm MeV}.
\end{eqnarray}
Adding this potential to the Hamiltonian in Eq.~(\ref{eq:Ham}),
our results agree with the observed energies of $^6$He for $0^+$ and $2^+$ states.
Hence, the present model reproduces the observed properties of $^{5,6}$He, simultaneously \cite{My07c,Aj89},
as shown in the next section (Fig.~\ref{fig:567}), namely, the threshold energies of the particle emissions from $^7$He.

\subsection{Complex scaling method (CSM)}

We explain the CSM, which describes resonances and nonresonant continuum states.
Hereafter, we refer to the nonresonant continuum states as simply the continuum states.
In the CSM, we transform the relative coordinates of the $^4$He+$Xn$ model shown in Fig.~\ref{fig:COSM}, by the operator $U_\theta$ as
\begin{eqnarray}
	U_\theta~:~~\vc{r}_i
&\to&	\vc{r}_i\, e^{i\theta} 
	\qquad \mbox{for}~~i=1,\cdots,X \ ,
\end{eqnarray}
where $\theta$ is a scaling angle.
The Hamiltonian in Eq.~(\ref{eq:Ham}) is transformed into the complex-scaled Hamiltonian $H_\theta=U_\theta H U_\theta^{-1}$, and the corresponding complex-scaled Schr\"odinger equation is given as
\begin{eqnarray}
	H_\theta\Psi^J_\theta
&=&     E\Psi^J_\theta,
	\label{eq:eigen}
	\\
        \Psi^J_\theta
&=&	e^{(3/2)i\theta\cdot X}\,
	\Psi^J(\{\vc{r}_i e^{i\theta}\}).
\end{eqnarray}
The eigenstates $\Psi^J_\theta$ are obtained by solving the eigenvalue problem of $H_\theta$ in Eq.~(\ref{eq:eigen}).
In CSM, we obtain all the energy eigenvalues $E$ of bound and unbound states on a complex energy plane, governed by the ABC theorem \cite{ABC}.
In this theorem, it is proven that the boundary condition of Gamow resonances 
is transformed to the damping behavior at the asymptotic region.
This condition enables us to use the same theoretical method to obtain the many-body resonances as that used for the bound states. 
For a finite value of $\theta$, every Riemann branch cut is commonly rotated down by $2\theta$.
For $^7$He, the continuum states of $^6${He}+$n$ $^5$He+2$n$ and $^4$He+3$n$ channels are obtained on the branch cuts rotated with the $2\theta$ dependence \cite{My07c}. On the contrary, bound states and resonances are discrete and obtainable independently of $\theta$ (see Fig. \ref{fig:ene} for the $^6$He($2^+$) case). 
Hence, these discrete states are located separately from the many-body continuum spectra on the complex energy plane.
We can identify the resonance poles of complex eigenvalues: $E=E_r-i\Gamma/2$, where $E_r$ and $\Gamma$ are the resonance energies measured from the lowest threshold and the decay widths, respectively. 

In the wave function, the $\theta$ dependence is included in the variational coefficients 
in Eqs.~(\ref{WF3}) and (\ref{WFR}) as $\{C_c^{J,\theta}\}$ and $\{C_{\alpha,k}^\theta\}$, respectively. 
We take the value of $\theta$ as 30$^\circ$ in this calculation.

In the study, we calculate the one-neutron removal strength of $^7$He into $^6$He,
in which we need a complete set of $^6$He including bound, resonant, and continuum states.
We express this $^6$He complete set using the complex-scaled eigenstates $\Psi^J_\theta$ obtained in the $^4$He+$n$+$n$ model.
Hereafter, we denote the $^6$He wave function with the state $\nu'$ as simply $\Phi_{\nu'}$,
and the $^7$He wave function $\Psi_\nu$($^7$He) as $\Psi_\nu$.

We briefly explain the extended completeness relation (ECR) of $^6$He using CSM \cite{My01,My98,Be68}.
When we take a large $\theta$ like in Fig.~\ref{fig:ene},
three-body unbound states of $^6$He is decomposed into three categories of discrete three-body resonances, three-body continuum states of $^4$He+$n$+$n$, and two-body continuum states of $^5$He(3/2$^-$, 1/2$^-$)+$n$.
Here, the $^5$He+$n$ two-body continuum states are obtained on the branch cuts, 
whose origins are resonance positions of $^5$He($3/2^-,1/2^-$), as shown in Fig.~\ref{fig:ene}.
Using all the unbound states of $^6$He, we introduce the extended three-body completeness relation (ECR) of the complex-scaled Hamiltonian $H_\theta$ of $^6$He as
\begin{eqnarray}
        {\bf 1}
&=&     \sum_\nu\hspace*{-0.5cm}\int\kets{\Phi^\theta_\nu}\bras{\wtil{\Phi}^\theta_\nu}
        \nonumber
        \\
&=&     \{\mbox{three-body bound state of $^6$He}\}
        \nonumber
        \\
&+&     \{\mbox{three-body resonance of $^6$He }\}
        \nonumber
        \\
&+&    \{\mbox{three-body continuum states of $^4$He+$n$+$n$}\}
        \nonumber
        \\
&+&     \{\mbox{two-body continuum states of $^5$He+$n$}\}\, ,
        \label{eq:3-ECR}
\end{eqnarray}
where $\{ \Phi_\nu^\theta,\wtil{\Phi}_\nu^\theta \}$ are the complex-scaled $^6$He wave functions 
and form a set of biorthogonal bases.
This relation is an extension of the two-body ECR \cite{Ao06,My98}.
Because the detailed definition of the biorthogonal bases is written in the previous works \cite{My01,My98}, 
we briefly explain it here.
When the wave number $k_\nu$ of $\Phi_\nu$ is for discrete bound and resonance states,
the adjoint wave number $\wtil{k}_\nu$ of $\wtil{\Phi}_\nu$ is defined as $\wtil{k}_\nu=-k^*_\nu$, 
which leads to the relation $\wtil{\Phi}_\nu$ = $(\Phi_\nu)^*$ \cite{My98,Be68,Mo78}. 
For continuum states, the same relation of the biorthogonal states of resonances is adopted, because we use a discretized representation.
In the $^4$He+$n$+$n$ model, the $^4$He-2$n$ channel is included in the three-body continuum components of $^4$He+$n$+$n$,
because $2n$ does not have any bound states or physical resonances.

\subsection{Spectroscopic factor of $^7$He}

We explain the $S$ factors of the $^6$He-$n$ components for $^7$He.
This $S$ factor is used in the calculation of the one-neutron removal strength of $^7$He.
As was explained in a previous study \cite{My07c}, because Gamow states generally give complex matrix elements,
$S$ factors of Gamow states are not necessarily positive definite and are defined by the squared matrix elements
using the biorthogonal property \cite{Mi07,My01,My07c,Be68} as
\begin{eqnarray}
    S^{J,\nu}_{J',\nu'}
&=& \sum_\alpha S^{J,\nu}_{J',\nu',\alpha}\, ,
\\
    S^{J,\nu}_{J',\nu',\alpha}
&=& \frac{1}{2J+1} \langle \widetilde{\Phi}^{J'}_{\nu'}||a_\alpha ||\Psi^J_\nu \rangle^2\, ,
    \label{eq:S}
\end{eqnarray}
where the annihilation operator $a_\alpha$ is for valence neutron with the state $\alpha$. 
The spin $J$ and $J'$ are for $^7$He and $^6$He, respectively.
The index $\nu$ ($\nu'$) indicates the eigenstate of $^7$He ($^6$He).
In this expression, the $S$ factors $S^{J,\nu}_{J',\nu'}$ are allowed to be complex values.
In general, an imaginary part of $S$ factor often becomes large relative to the real part for resonances having large decay widths.

The sum rule value of $S$ factors, which includes Gamow state contributions of the final states, can be considered \cite{My07c,Ao06}.
When we count all the obtained complex $S$ factors not only of the Gamow states but also of the continuum states in the final states,
the summed value of $S$ factors satisfies the associated particle number, which is a real value.
In such a case, the imaginary part of the summed $S$ factors becomes zero, similar to the transition strength \cite{Ao06,My01,My03}.
For $^7$He into the $^6$He-$n$ decomposition, the summed value of the $S$ factors $S^{J,\nu}_{J',\nu'}$ in Eq.~(\ref{eq:S}) by taking all the $^6$He states, is given as
\begin{eqnarray}
    \sum_{J',\nu'}\ S^{J,\nu}_{J',\nu'}
&=& \sum_{\alpha,m}\ 
    \langle \widetilde{\Psi}^{JM}_\nu|a^\dagger_{\alpha,m} a_{\alpha,m}| \Psi^{JM}_\nu \rangle
    \nonumber
    \\
&=& 3\ ,
    \label{eq:sf-sum}
\end{eqnarray} 
where we use the completeness relation of $^6$He as
\begin{eqnarray}
    1
&=& \sum_{J',M'}\sum_{\nu'}\hspace*{-0.5cm}\int |\Phi^{J'M'}_{\nu'}\rangle \langle \widetilde{\Phi}^{J'M'}_{\nu'}|.
\end{eqnarray}
Here the labels, $M$ ($M'$) and $m$ are the $z$ components of the angular momenta of the wave functions of $^7$He ($^6$He) and 
of the creation and annihilation operators of valence neutrons, respectively.
It is found that the summed value of $S$ factors for the $^6$He states becomes the number of valence neutrons in $^7$He.
This property of the $S$ factors is also established when the complex scaling is operated.

In the numerical calculation, we express the radial part of the operator $a_\alpha$ in Eq.~(\ref{eq:S}) using a complete set expanded by 
40 Gaussian basis functions with the maximum range of 80 fm for each orbit. 
This treatment is sufficient to converge the results.

\subsection{One-neutron removal strength of $^7$He}\label{sec:spec}

We explain the one-neutron removal strength of $^7$He into $^6$He.
This is a function of the real energy of $^6$He, $E$.
We first introduce the Green's function ${\cal G}(E,\vc{\eta},\vc{\eta}')$ of $^6$He,
which is used in the derivation of the strength.
The procedure is the same as that used in the case of the electric transitions \cite{My01,My03}.
The coordinates, $\vc{\eta}$ and $\vc{\eta}'$, represent the set of $\vc{r}_i$ ($i=1,\cdots,X$) in Fig.~\ref{fig:COSM}.
Here, we introduce the complex-scaled Green's function ${\cal G}^\theta(E,\vc{\eta},\vc{\eta}')$ of $^6$He as
\begin{eqnarray}
        {\cal G}(E,\vc{\eta},\vc{\eta}')
&=&     \left\bra \vc{\eta} \left|
        \frac{ {\bf 1} }{ E-H }\right|\vc{\eta}' \right\ket
        \label{eq:green0}
        \\
~\to~   {\cal G}^\theta(E,\vc{\eta},\vc{\eta}')
&=&     \left\bra \vc{\eta} \left|
        \frac{ {\bf 1} }{ E-H_\theta }\right|\vc{\eta}' \right\ket
        \nonumber
        \\
&=&     \sum_\nu\hspace*{-0.5cm}\int\
        \frac{\Phi^\theta_\nu(\vc{\eta})\ [\wtil{\Phi}^*_\nu(\vc{\eta}')]^\theta}{E-E_\nu^\theta}
        \nonumber
        \\
&=&     \sum_\nu\hspace*{-0.5cm}\int\
        {\cal G}^\theta_\nu(E,\vc{\eta},\vc{\eta}')\, .
        \label{eq:green1}
\end{eqnarray}
In the derivation from Eq.~(\ref{eq:green0}) to Eq.~(\ref{eq:green1}), we insert the ECR of $^6$He given in Eq.~(\ref{eq:3-ECR}). 
The $^6$He energy, $E_\nu^\theta$, corresponds to the eigen wave function $\Phi^\theta_\nu$.
The $\theta$ dependence of $E_\nu^\theta$ appears only in the continuum spectra. 

Next, the strength function $S(E)$ for the annihilation operator $a_\alpha$ 
is defined using the Green's function in a usual case without CSM as
\begin{eqnarray}
        S(E) 
&=&     \sum_\alpha S_\alpha(E) ,
        \label{eq:spec}
        \\
        S_\alpha(E) 
&=&     \sum_\nu \hspace*{-0.5cm}\int\
        \bras{\wtil{\Psi}_0}a^\dagger_\alpha \kets{\Phi_\nu}\bras{\wtil{\Phi}_\nu} a_\alpha \kets{\Psi_0}\
        \delta(E-E_\nu)
        \label{eq:strength_org}
        \nonumber
        \\
&=&     -\frac1{\pi}\ {\rm Im}
        \left[
        \int d\vc{\eta} d\vc{\eta}'                    \:
        \wtil{\Psi}_0^*(\vc{\eta})\: a^\dagger_\alpha  \:
        \right.
        \nonumber
        \\
&\times&{\cal G}(E,\vc{\eta},\vc{\eta}')               \:
        a_\alpha \Psi_0(\vc{\eta}')
        \biggr]\, .
        \label{eq:strength0}
\end{eqnarray}
For simplicity, we omit the labels of the angular momenta and their $z$ components of the wave functions and of the operators. 
The wave function $\Psi_0$ is the ground state of $^7$He.
We also consider the sum rule value of the strength $S(E)$ in Eq.~(\ref{eq:spec}),
which is defined by the integration of $S(E)$ over the real energy $E$ of $^6$He.
Using the completeness relation of the final states of $^6$He, the sum rule value is given as
\begin{eqnarray}
        \int dE\ S(E) 
&=&     \sum_{\alpha} \sum_{\nu}\hspace*{-0.5cm}\int\ \bras{\wtil{\Psi}_0}a^\dagger_\alpha \kets{\Phi_\nu}\bras{\wtil{\Phi}_\nu} a_\alpha \kets{\Psi_0}\
        \nonumber
        \\
&=&     \sum_{\alpha} \langle \widetilde{\Psi}_0|a^\dagger_{\alpha} a_{\alpha}| \Psi_0 \rangle
        \nonumber
        \\
&=&     3\, .
\end{eqnarray}
Thus, it is also confirmed that the energy integrated value of $S(E)$ satisfies the associated particle number of $^7$He, similar to the case shown in Eq.~(\ref{eq:sf-sum}).

To calculate the strength function $S_\alpha(E)$ in Eq.~(\ref{eq:strength0}), 
we operate the complex scaling on $S_\alpha(E)$, 
and use the complex-scaled Green's function of Eq.~(\ref{eq:green1}) as
\begin{eqnarray}
        S_\alpha(E)
&=&     -\frac1{\pi}\ {\rm Im}
        \left[
        \int d\vc{\eta} d\vc{\eta}'                       \:
        [\wtil{\Psi}_0^*(\vc{\eta})]^\theta (a^\dagger_\alpha)^\theta\:
        \right.
        \nonumber
         \\
&\times&{\cal G}^\theta(E,\vc{\eta},\vc{\eta}')            \:
        a_\alpha^\theta \Psi^\theta_0(\vc{\eta}')
        \biggr]
        \nonumber
        \\
&=&     \sum_\nu\hspace*{-0.5cm}\int\ S_{\alpha,\nu}(E)\, ,
        \label{eq:strength1}
        \\
        S_{\alpha,\nu}(E)
&=&     -\frac1{\pi}\ {\rm Im}
        \left[   \frac{
                \bras{\wtil{\Psi}_0^\theta}  (a^\dagger_\alpha)^\theta \kets{\Phi_\nu^\theta}
                \bras{\wtil{\Phi}_\nu^\theta} a_\alpha^\theta          \kets{\Psi_0^\theta}
          }{E-E_\nu^\theta}\right] .
        \label{eq:strength2}
\end{eqnarray}
In Eq.~(\ref{eq:strength2}), the strength function is calculated using the one-neutron removal matrix elements
$\bras{\wtil{\Phi}_\nu^\theta} a_\alpha^\theta\kets{\Psi^\theta_0}$.
It is noted that the function $S_{\alpha,\nu}(E)$ is independent of $\theta$ \cite{My07c,My01,My98,Su05}.
This is because any matrix elements are obtained independently of $\theta$ in the complex scaling method, 
and also because the state $\nu$ of $^6$He is uniquely classified according to ECR defined in Eq.~(\ref{eq:3-ECR}).
As a result, the decomposed strength $S_{\alpha,\nu}(E)$ is uniquely obtained.
Thus, the one-neutron removal strength $S_\alpha(E)$ is calculated as a function of the real energy $E$ of $^6$He.
When we discuss the structures appearing in $S_\alpha(E)$, it is useful to decompose $S_\alpha(E)$
into each component $S_{\alpha,\nu}(E)$ by using the complete set of the final state $\nu$ of $^6$He.
We can categorize $\nu$ of $^6$He using the ECR in Eq.~(\ref{eq:3-ECR}).
Because of this decomposition of unbound states,
we can unambiguously investigate how much each resonant and continuum state of $^6$He exhausts the strength.
This is a prominent point of the present method.

Here, we discuss the properties of the decomposed strength function $S_{\alpha,\nu}(E)$ in Eq.~(\ref{eq:strength2}).
The original strength function $S_{\alpha}(E)$ in Eq.~(\ref{eq:strength1}) corresponds to an observable being positive definite for every energy. 
However, the decomposed strength function $S_{\alpha,\nu}(E)$ is not necessarily positive definite at all energies,
because $S_{\alpha,\nu}(E)$ cannot be directly observed, similar to the resonant poles.
This means that $S_{\alpha,\nu}(E)$ can sometimes have negative values.
This property of the decomposed strength has been generally discussed in the electric transition strength \cite{My01,My03}.

The stability of the calculated matrix elements of resonant and continuum states using the CSM 
has been shown in many works \cite{Ao06,My01,My98}.
For continuum states, we adopt the discretized representation using the $L^2$ integrable basis functions.
This discretization has been checked to reproduce the genuine continuum states by using the CSM \cite{Kr07,Su08,Su05}.

\section{Results}\label{sec:result}

\subsection{Energy spectra of He isotopes}

\begin{figure}[b]
\centering
\includegraphics[width=8.0cm,clip]{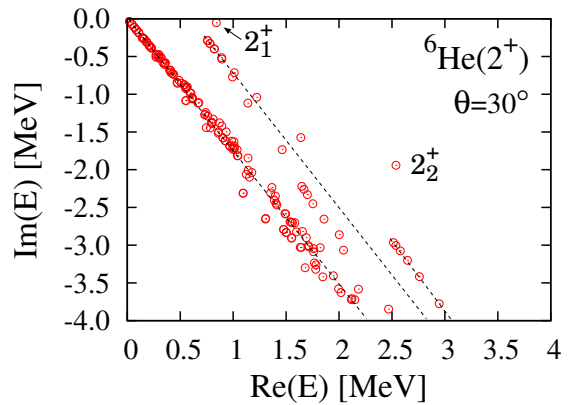}
\caption{(Color online) Energy eigenvalues of the $^6$He($2^+$) states in the complex energy plane
measured from the $^4$He+$n$+$n$ threshold. 
Three schematic lines with dots correspond to the $^4$He+$n$+$n$ and $^5$He($3/2^-$,$1/2^-$)+$n$ continuum states,
in order from left to right, respectively.}
\label{fig:ene}n
\end{figure}

\begin{figure}[th]
\centering
\includegraphics[width=8.0cm,clip]{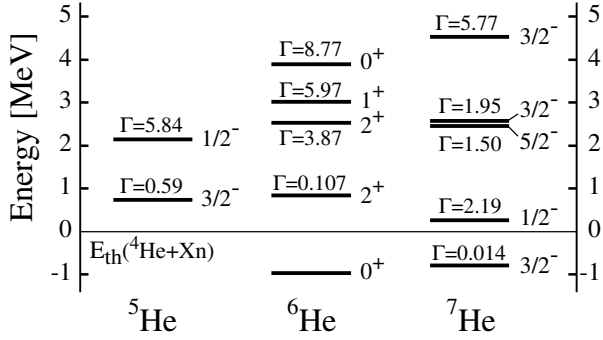}
\caption{Energy eigenvalues of the obtained $^{5,6,7}$He resonances measured from the $^4$He+$Xn$ threshold.}
\label{fig:567}
\end{figure}

We first discuss the energy spectra of $^{5-7}$He obtained in the present model.
The example of the eigenvalue distribution of $^6$He($2^+$) obtained in the CSM is shown in Fig.\ref{fig:ene}.
In this figure, we can easily identify the locations of the three-body resonances of $^6$He ($2^+_1$ and $2^+_2$) and,
further, of the two kinds of continuum states of $^5$He($3/2^-$,$1/2^-$)+$n$ and $^4$He+$n$+$n$.
In Fig.~\ref{fig:567}, we summarize the energy spectra of $^{5-7}$He.
A detailed discussion of the obtained energy levels is given in previous works \cite{My07c,Ao06};
and hence we briefly explain it here.

In Tables~\ref{ene6} and \ref{ene7}, we list the energies and decay widths of $^6$He and $^7$He, respectively, 
measured from the $^4$He+$Xn$ threshold energy.
For $^7$He, the ground state energy $E_r$ is 0.184 MeV measured from the $^6$He+$n$ threshold, 
which is slightly overbound with respect to the experiments ($E_r=0.44(2)$ MeV  \cite{Ko99} and $0.36(5)$ MeV  \cite{Sk06}).
Due to this overbinding, the calculated decay width is smaller than the experiments of $\Gamma \sim 0.16$ MeV \cite{Ko99,Sk06}. 
When we fit the observed energy of $E_r=0.44$ MeV, namely, $-$0.54 MeV from the $^4$He+$3n$ threshold, by reducing the strength of $V^{\alpha n n}$,
the decay width becomes 0.14 MeV and nicely agrees with the experiments, as is shown in Table~\ref{ene7}.

\begin{table}[th]
\caption{Energy eigenvalues of the $^{6}$He resonances measured from the $^4$He+2$n$ threshold.
The values with parentheses are the experimental ones \cite{Aj89}. Dominant configurations are also listed.}
\label{ene6}
\centering
\begin{ruledtabular}
\begin{tabular}{c|cccc}
          & Energy~(MeV)         &  Width~(MeV)      & Config.              \\ \hline
 $0^+_1$  & $-0.974$~($-0.975$)  &  ---              & $(p_{3/2})^2$        \\
 $0^+_2$  & $3.90$               &  8.77             & $(p_{1/2})^2$        \\
 $2^+_1$  & $0.840$~(0.822(25))  &  0.107~(0.113(20))& $(p_{3/2})^2$        \\
 $2^+_2$  & $2.53$               &  3.87             & $(p_{3/2})(p_{1/2})$ \\
 $1^+$    & $3.02$               &  5.97             & $(p_{3/2})(p_{1/2})$ \\
\end{tabular}
\end{ruledtabular}
\end{table}

\begin{table}[th]
\caption{Energy eigenvalues of the $^7$He resonances measured from the $^4$He+3$n$ threshold.
The values with parentheses are the ones fitted to the position of the observed resonance energy of the ground state.
Dominant configurations are also listed.}
\label{ene7}
\centering
\begin{ruledtabular}
\begin{tabular}{c|cccc}
            & Energy~(MeV)        &  Width~(MeV) & Config.                \\ \hline
 $3/2^-_1$  & $-0.790$~($-$0.54)  &  0.014~(0.14)& $(p_{3/2})^3$          \\
 $3/2^-_2$  & $2.58$              &  1.95        & $(p_{3/2})^2(p_{1/2})$ \\
 $3/2^-_3$  & $4.53$              &  5.77        & $(p_{3/2})(p_{1/2})^2$ \\
 $1/2^-  $  & $0.26$              &  2.19        & $(p_{3/2})^2(p_{1/2})$ \\
 $5/2^-  $  & $2.46$              &  1.50        & $(p_{3/2})^2(p_{1/2})$ \\
\end{tabular}
\end{ruledtabular}
\end{table}

\begin{table}[t]
\caption{Spectroscopic factors of the $^6$He-$n$ components of $^7$He. Details are described in the text.}
\label{sf}
\begin{ruledtabular}
\begin{tabular}{c|ll|ll}
          & \multicolumn{2}{c|}{$^6$He($0^+_1$)-$n$}     &  \multicolumn{2}{c}{$^6$He($2^+_1$)-$n$} \\
          & Present        & CK   &  Present      & CK    \\ \hline
$3/2^-_1$ & $0.64 +i0.06$  & 0.59 &  $1.55-i0.31$ & 1.21  \\
$3/2^-_2$ & $0.005+i0.01$  & 0.06 &  $0.95+i0.02$ & 1.38  \\
$3/2^-_3$ & $0.003+i0.0002$& ---  &  $0.02-i0.004$& ---   \\ 
$1/2^- $  & $1.00 -i0.13$  & 0.69 &  $0.10-i0.05$ & 0.60  \\
$5/2^- $  & $0.00+i0.00$   & 0.00 &  $0.95+i0.02$ & 1.36  \\
\end{tabular}
\end{ruledtabular}
\end{table}

\begin{table}[t]
\caption{Components of two $p$ orbits of the last neutron in $S$ factors of $^6$He-$n$($p$-wave) of $^7$He. 
Upper and lower tables correspond to $^6$He($0^+_1$) and $^6$He($2^+_1$), respectively.}
\label{sf2}
\begin{ruledtabular}
\begin{tabular}{c|ccc}
          & $^6$He($0^+_1$)-$p_{3/2}$       & $^6$He($0^+_1$)-$p_{1/2}$     \\
$3/2^-_1$ & $0.64 +i0.06$   & $0.00+i0.00$  \\
$3/2^-_2$ & $0.005+i0.01$   & $0.00+i0.00$  \\
$3/2^-_3$ & $0.003+i0.0002$ & $0.00+i0.00$  \\ 
$1/2^- $  & $0.00+i0.00$    & $1.00-i0.13$  \\
$5/2^- $  & $0.00+i0.00$    & $0.00+i0.00$  \\ \hline
\\ 
          & $^6$He($2^+_1$)-$p_{3/2}$       & $^6$He($2^+_1$)-$p_{1/2}$     \\
$3/2^-_1$ & $1.54 -i0.31$   & $0.005-i0.002$ \\
$3/2^-_2$ & $0.001+i0.001$  & $0.95+i0.02$   \\
$3/2^-_3$ & $0.007-i0.015$  & $0.02+i0.01$   \\ 
$1/2^- $  & $0.10-i0.05$    & $0.00+i0.00$   \\
$5/2^- $  & $0.10+i0.02$    & $0.85-i0.001$  \\
\end{tabular}
\end{ruledtabular}
\end{table}

We found the $5/2^-$ state, whose position agrees with the several experiments \cite{Ko99,Bo01,Sk06}.
Further, the $3/2^-_2$ state is degenerated with the $5/2^-$ state and their decay widths do not differ so much.
This result suggests the possibility of the superposed observation of two states in this energy region.
We found one broad $1/2^-$ resonance at a low excitation energy of $E_x$=1.05 MeV,
while the experimental uncertainty is still large \cite{Me02,Sk06,Ry06,Wu05,Bo05,Wu08}.
Recently, there has been a report discussing the order of energy levels \cite{Wu08}, which fully agrees with our results.

For the neutron configurations, the resonances of $^6$He and $^7$He are dominantly described 
by the $p$-shell configurations and the small contributions come from the $sd$-shell \cite{My07c}.
The dominant configurations for each resonance are listed in Tables~\ref{ene6} for $^6$He and Table \ref{ene7} for $^7$He, respectively. 
From the configurations, several excited states of $^7$He seem to have the $^6$He+$n$ type configurations.
To understand the detailed structures of the $^7$He states, we investigate 
the $S$ factors of the $^6$He+$n$ system and the one-neutron removal strength of $^7$He.

\subsection{Spectroscopic factors of $^7$He}\label{sec:sfac}

We calculate the $S$ factors of the $^6$He-$n$ components of the $^7$He resonances.
Here we choose the $0^+_1$ and $2^+_1$ states of $^6$He.
It is noted that the present $S$ factors correspond to the components of $^6$He inside the $^7$He resonances and contain the imaginary part.
We improve the operation of complex scaling to calculate the matrix elements of $S$ factors from the previous ones \cite{My07c}. 
In Table~\ref{sf}, we list the results of $S$ factors of each $^7$He resonance. 
For reference, the results of the conventional Cohen-Kurath shell model (CK) \cite{Wu05} are shown
with real values due to the bound state approximation of resonances.
With regard to the real parts of the calculated $S$ factors, the trend seen in our results is roughly similar to that of the CK results.
In Table~\ref{sf2}, the $S$ factors are decomposed into the components of two $p$ orbits of the last neutron.
It is found that every $^7$He resonance is dominated by one of the $p$ orbits coupled with $^6$He.

In Table~\ref{sf}, for the $3/2^-_1$ state, the $^6$He($0^+_1$)-$n$ component almost shows a real value with a small imaginary part.
This real part well corresponds to the recent observation of $0.64\pm0.09$  \cite{Be07}.
The $^6$He($2^+_1$)-$n$ component is large, more than twice of that of the $^6$He($0^+_1$)-$n$ case.
For the $3/2^-_2$ state, $^6$He($2^+_1$) is dominantly mixed, and this characteristic is common for the $5/2^-$ state.
Hence, these two states have a similar structure of the valence neutron configurations.
For the $3/2^-_3$ state, two components of $^6$He($0^+_1$, $2^+_1$)-$n$ are very small.
This state can be considered to have some components of $^6$He($0^+_2$,$1^+$,$2^+_2$)-$n$ 
from the configurations shown in Tables \ref{ene6} and \ref{ene7}.

For the $1/2^-$ state, the $S$ factor of $^6$He($0^+_1$)-$n(p_{1/2})$ is almost unity with a small imaginary part, 
and the $^6$He($2^+_1$)-$n$ component is small.
These results of $1/2^-$ indicate that the $^6$He($0^+_1$)-$n$ component is dominant in $^7$He($1/2^-$)
and also suggest the weak coupling nature of the $p_{1/2}$ orbital neutron around $^6$He, 
which retains two-neutron halo structure.
Therefore, the $1/2^-$ resonance could be mainly considered as a single particle resonance of the $p_{1/2}$ neutron surrounding a halo state of $^6$He.

We obtained the interesting information of the structures of the $^7$He resonances via $S$ factors.  
We should be, however, careful to derive the conclusion of physical interpretation of the complex $S$ factors of Gamow states,
as was mentioned in the previous studies \cite{My07c,My01,My03}.
In this analysis, both of the initial ($^7$He) and final ($^6$He) states are Gamow states. 
In the next subsection, we discuss a one-neutron removal strength from the $^7$He ground state assuming a bound state wave function 
to the continuum energy states of $^6$He. 
The continuum energy states of $^6$He are described using the complex-scaled complete set of $^6$He.
It is shown that we can see contributions not only of resonances but also of nonresonant continuum states 
of $^6$He to the strength function.

\subsection{One-neutron removal strength of $^7$He}

In Sec.~\ref{sec:sfac}, the obtained $S$ factors are complex values, which are caused 
from the Gamow states of $^7$He and $^6$He($2^+_1$).
For the final states of $^6$He, we consider only resonances, not the continuum states of $^5$He+$n$ and $^4$He+$n$+$n$.
The decay contributions from $^7$He to these continuum states are not clarified.
In this study, we take up this problem and show the strength function of the $^7$He ground state into $^6$He 
using Eqs.~(\ref{eq:strength1}) and (\ref{eq:strength2}).
For $^6$He, we prepare the three-body complete set of the $^4$He+$n$+$n$ model, namely, ECR in Eq.~(\ref{eq:3-ECR})
by using the complex-scaled wave functions $\Phi_\nu^\theta$ of $^6$He.
In this method, we take care not only of the $^6$He resonances, but also of the continuum states of $^5$He+$n$ and $^4$He+$n$+$n$.
Here, we use the bound state approximation for the initial $^7$He ground state,
because this state has a very small decay width, as shown in Table~\ref{ene7}.

\begin{figure}[t]
\centering
\includegraphics[width=8.0cm,clip]{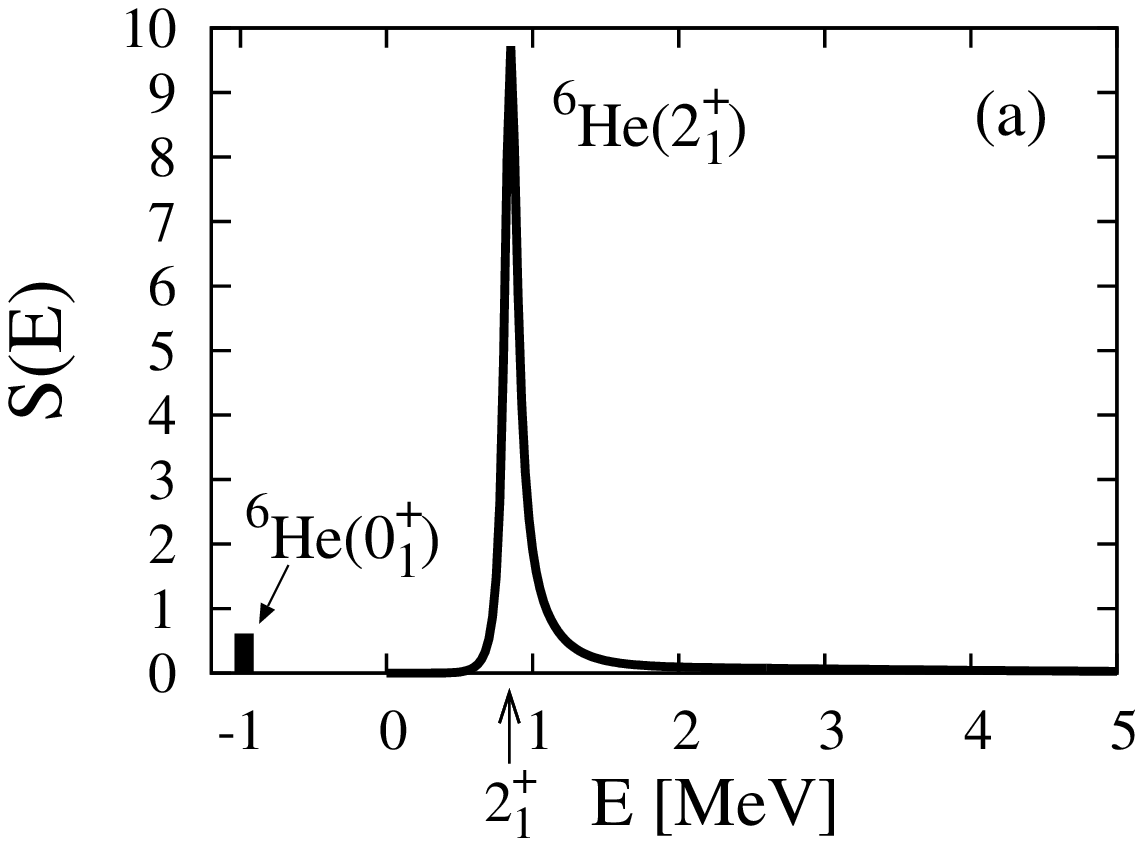}
\includegraphics[width=8.0cm,clip]{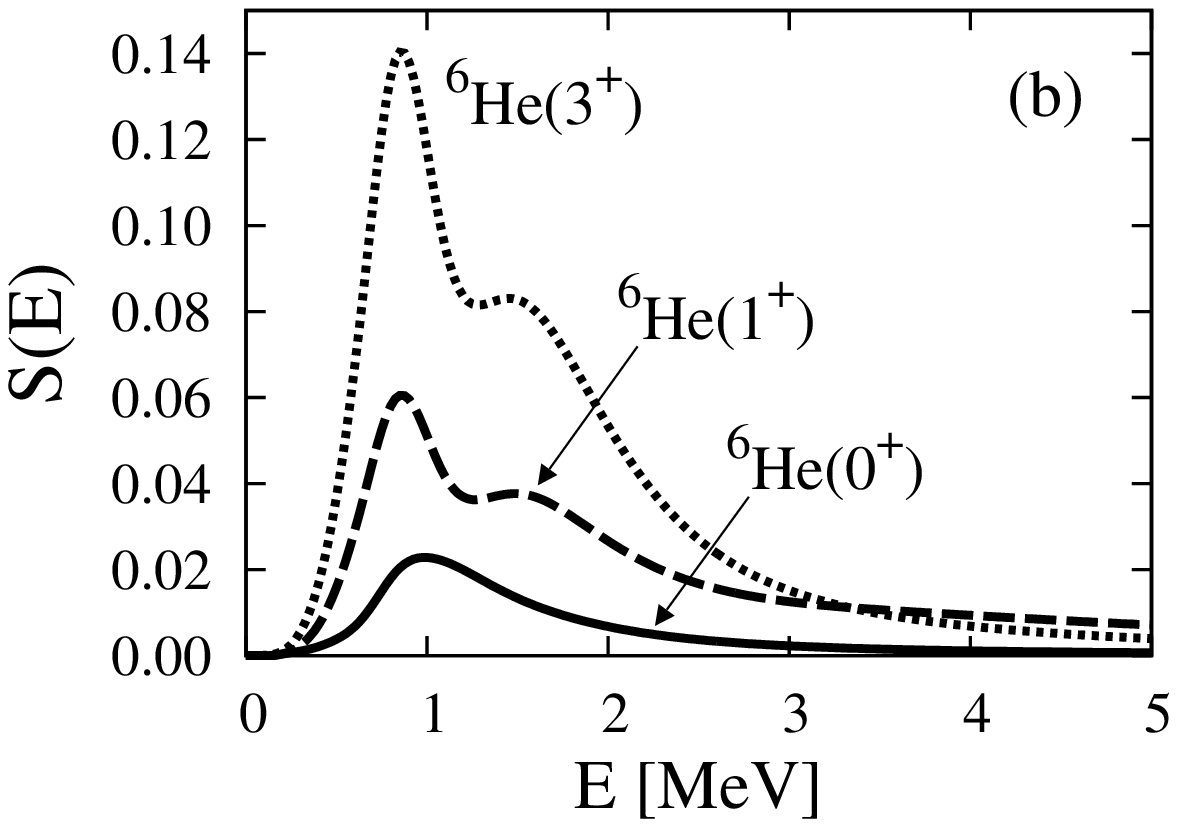}
\caption{One-neutron removal strength of $^7$He into $^6$He($J^\pi$,$E$) states measured from the $^4$He+$n$+$n$ threshold energy.
The vertical arrow in (a) indicates the resonance energy of $^6$He($2^+_1$).
The strength into $^6$He($0^+_1$) is shown by a histogram at the position of the corresponding energy.}
\label{fig:spec}
\end{figure}

\begin{figure}[b]
\centering
\includegraphics[width=8.0cm,clip]{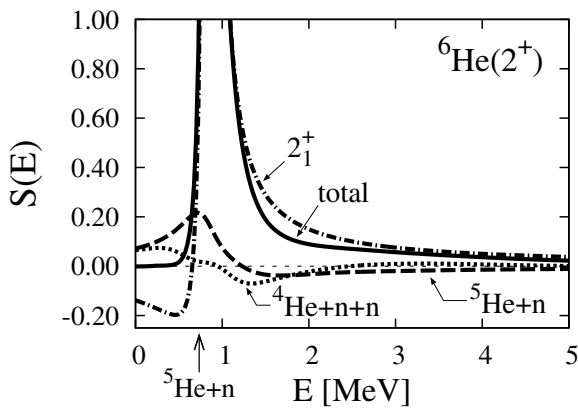}
\caption{Decomposition of the $^6$He($2^+$) component of the strength functions.
The vertical arrow indicates the threshold energy (0.74 MeV) of the $^5$He($3/2^-$)+$n$ channel.}
\label{fig:sep1}
\end{figure}

\begin{figure}[hb]
\centering
\includegraphics[width=8.0cm,clip]{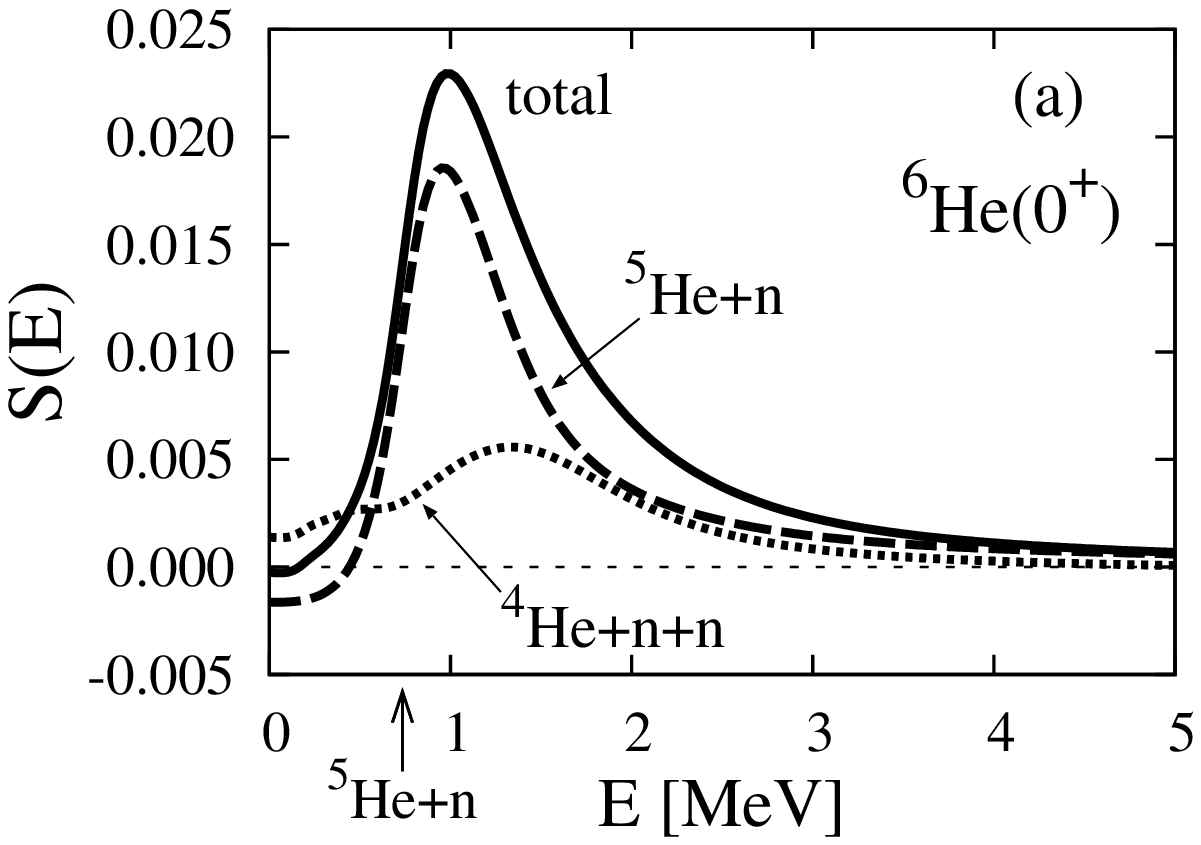}
\includegraphics[width=8.0cm,clip]{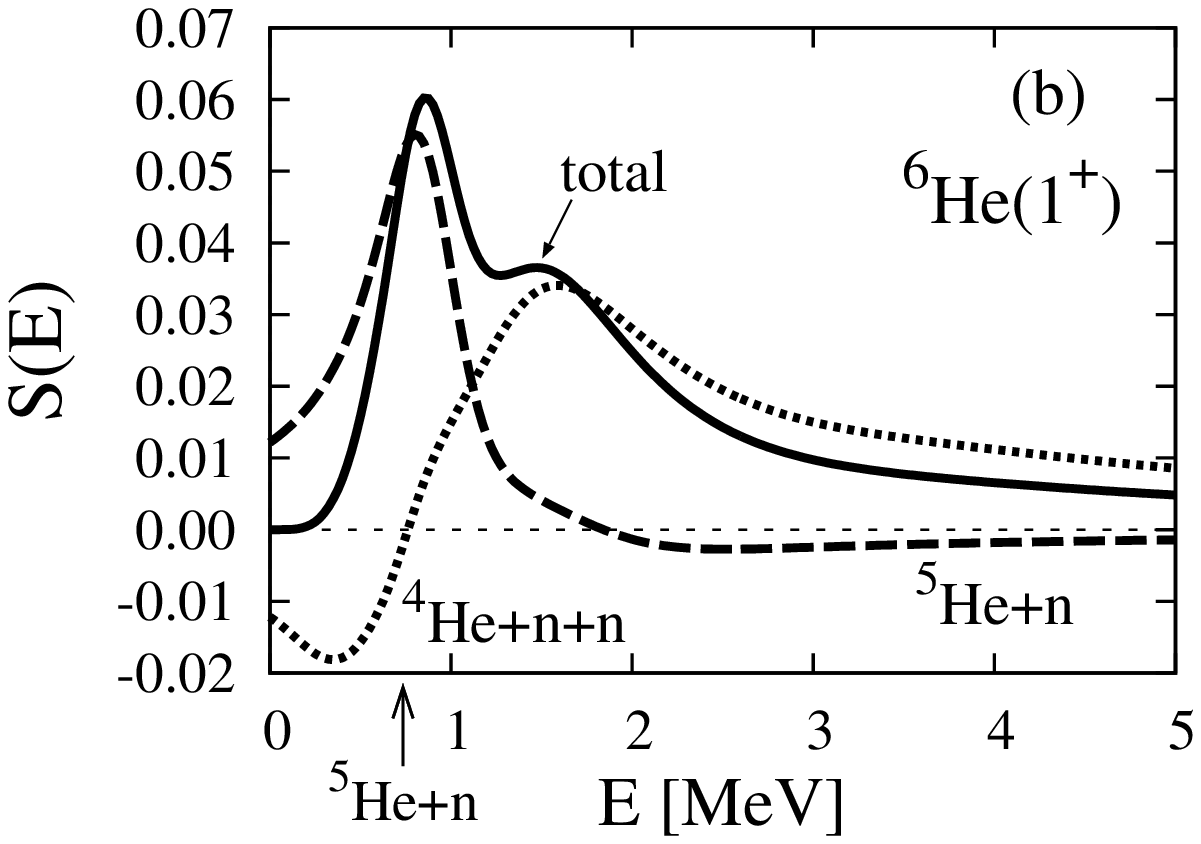}
\includegraphics[width=8.0cm,clip]{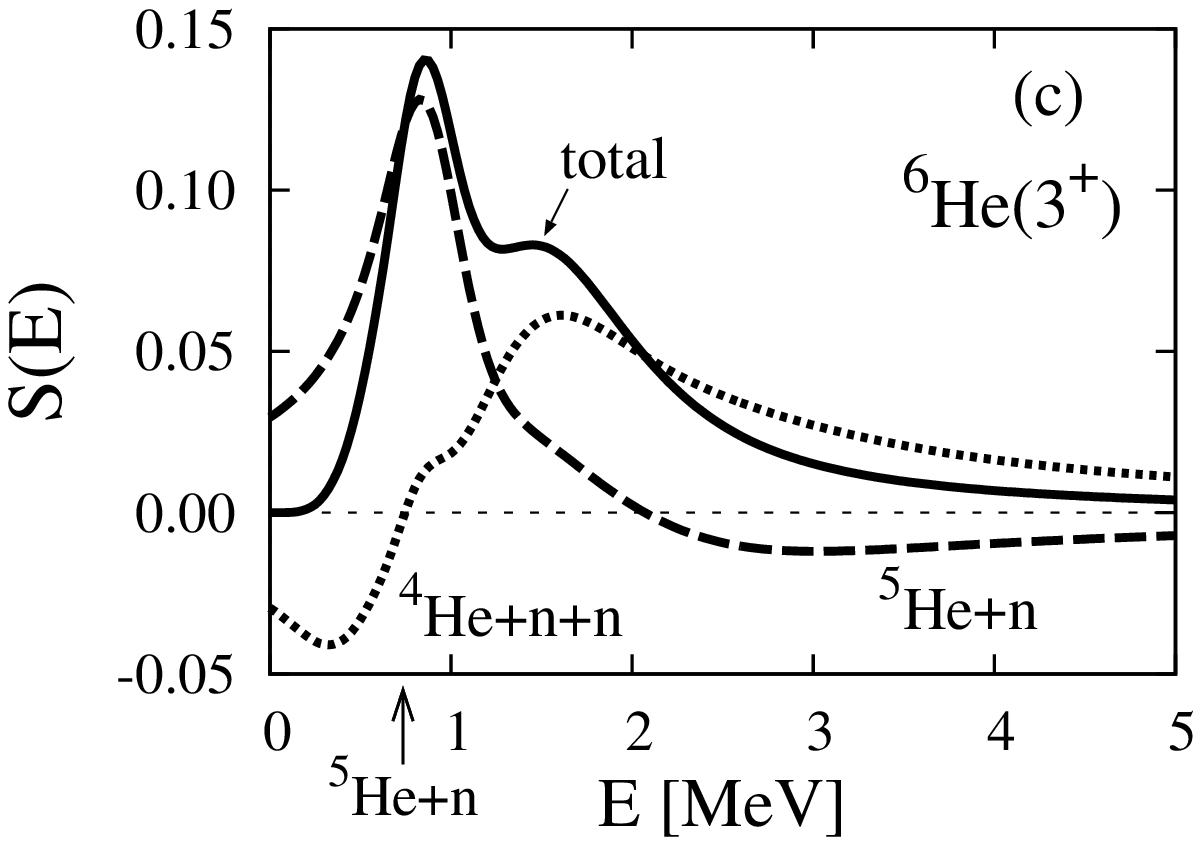}
\caption{Decomposition of the $^6$He($0^+$,$1^+$,$3^+$) components of the strength functions.
The vertical arrows in each panel indicate the threshold energy of the $^5$He($3/2^-$)+$n$ channel.}
\label{fig:sep2}
\end{figure}

In Fig.~\ref{fig:spec}, we show the one-neutron removal strength $S(E)$ of $^7$He$(3/2^-)$ into $^6$He with spin $J$, 
where the energy $E$ of $^6$He is measured from the $^4$He+$n$+$n$ threshold.
We show the results of only plus parity states of $^6$He, 
while the negative parity strengths are found to give a negligible contribution in the order of $10^{-3}$ of $S(E)$.
In Fig.~\ref{fig:spec}, the obtained strengths remain positive definite.
The dominant component comes from the $2^+$ state in Fig.~\ref{fig:spec} (a), whose strength has a peak at the resonance energy, 0.84 MeV,
of $^6$He($2^+_1$).
Below the $^4$He+$n$+$n$ threshold, $^6$He($0^+_1$) gives a contribution of 0.61 of $S(E)$,
which is close to the real part of 0.64 in Table \ref{sf}.
The small difference comes from the bound state approximation of the $^7$He ground state.
For the continuum energy region above the $^4$He+$n$+$n$ threshold, 
the $0^+$, $1^+$ and $3^+$ states give small contributions in the strengths, as shown in Fig.~\ref{fig:spec} (b). 
It is also found that three kinds of strengths commonly show the enhancement at around 1 MeV.
From these results, it is concluded that the one-neutron removal strength of $^7$He is dominantly exhausted 
by the $^6$He($2^+_1$) resonance above the $^4$He+$n$+$n$ threshold energy.

We discuss the detailed structures seen in the four strengths in Fig.~\ref{fig:spec}. 
The origin of the $2^+$ peak is the $^6$He($2^+_1$) resonance, 
whose matrix element $(1.55-i0.31)$ in Table \ref{sf} is so large that 
the strength function $S(E)$ produces the sharp peak due to the small decay width of 0.11 MeV of $^6$He($2^+_1$).
The remaining continuum strength of the $^5$He+$n$ and $^4$He+$n$+$n$ components are very small and masked by the sharp peak from resonance.
For the $1^+$ and $3^+$ states, we see one peak and a shoulder-like structure. For the $0^+$ state, only one peak is seen.
It is interesting to examine the origins of these structures in each strength.
We do this it by decomposing the strength functions using the ECR of $^6$He in Eq.~(\ref{eq:strength2}).

In Fig.~\ref{fig:sep1}, the $2^+$ strength distribution is decomposed into the three kinds of the components
of $^6$He($2^+_1$) resonance, $^5$He($3/2^-$)+$n$ and $^4$He+$n$+$n$ continuum states.
It is clearly and explicitly shown that the sharp peak is exhausted by the $^6$He($2^+_1$) resonance.
The residual $2^+$ continuum strengths are found to be small.
Among them, the $^5$He($3/2^-$)+$n$ two-body continuum component makes a peak at around 0.75 MeV. 
This energy coincides with the position of the $^5$He+$n$ threshold energy (0.74 MeV), and then
the peak reflects the threshold effect of the $^5$He+$n$ open channel.
The $^4$He+$n$+$n$ contribution is small and does not produce a definite structure in the distribution.

In Fig.~\ref{fig:sep2}, the strengths of $^6$He($0^+$,$1^+$,$3^+$) are decomposed into two kinds of continuum components.
In every figure, it is found that the low-energy peak observed commonly at around 1 MeV, 
comes from the contribution of $^5$He($3/2^-$)+$n$.
This peak can be considered as a threshold effect of the two-body $^5$He($3/2^-$)+$n$ channel.
The $^4$He+$n$+$n$ component also commonly makes a mild bump at around 1.5 MeV, 
which makes the shoulder-like structure in the total strength for $1^+$ and $3^+$, in Fig.~\ref{fig:sep2} (b) and (c),
respectively.
This $^4$He+$n$+$n$ strength is considered to correspond to the background contribution 
and does not indicate the existence of any physical states.
For the $0^+$ state in Fig.~\ref{fig:sep2} (a), the $^4$He+$n$+$n$ component is smaller than the $^5$He($3/2^-$)+$n$ one,
and then only one peak structure is confirmed in the total strength.
As a result, it is found that for three spin states, the strengths into the $^6$He unbound states show some structures, 
whose origins come from the two kinds of different continuum components of $^5$He+$n$ and $^4$He+$n$+$n$.

It is noted that there are other components of broad resonances such as $^6$He($0_2^+$, $1^+$, $2_2^+$) and 
also a binary $^5$He($1/2^-$)+$n$ one.
Their contributions are checked to give negligible contributions in the strength.
This is because the magnitudes of their matrix elements are very small, and further, these states have large decay widths as shown in Table~\ref{ene6}. 
For $^5$He($1/2^-$), its decay width is obtained as 5.84 MeV.
Similar results from broad resonances were confirmed in the Coulomb breakup strengths of halo nuclei \cite{My01,My03}.

It is found in Figs.~\ref{fig:sep1} and \ref{fig:sep2} that the $^5$He($3/2^-$)+$n$ two-body continuum component 
remains with finite values even below the two-body threshold energy.
This is because  the $^5$He in the $^5$He+$n$ threshold, has a decay width of 0.60 MeV.
Hence, the $^5$He+$n$ component of $S(E)$ can have a strength below the threshold energy \cite{My01}.

In this model, the $^7$He ground state energy is slightly overbound with respect to the experiments.
Hence, we also calculate the strengths when we fit the observed ground state energy of $^7$He, as shown in Table~\ref{ene7}.
It is confirmed that the trend of the obtained strengths does not depend on the $^7$He ground state energy,
and the present conclusion does not change. 

\section{Summary}\label{sec:summary}

We have investigated the structures of $^7$He with the four-body cluster model.
The boundary condition for many-body resonances is accurately treated using the complex scaling method. 
In this study, we have developed the analysis of $^7$He. 
From the $S$ factors of the $^6$He-$n$ component of $^7$He,
the $1/2^-$ resonance has a possibility to be the weakly coupled state consisting of 
a halo state of $^6$He and a $p_{1/2}$ valence neutron.

We have also proposed the method to obtain the real-value strength functions 
from Gamow states giving complex-value $S$ factors.
This leads to the calculation of the one-neutron removal strength of $^7$He into $^6$He.
We calculate this strength function by using the matrix elements of the resonant and the nonresonant continuum states obtained with the complex scaling method.
Using the complex-scaled complete set of $^6$He, we take into account the strength into the continuum states in addition to the resonances for the final states of $^6$He.

The one-neutron removal strength of $^7$He into $^6$He is successfully obtained as an observable with positive definite.
From the results, the importance of $^6$He($2^+_1$) is clearly shown. 
However, the many-body continuum components of $^5$He+$n$ and $^4$He+$n$+$n$ are found to give 
small contributions, although they show some structures, such as the threshold effect, in the distributions.

In the present analysis, we treat only the $^7$He ground state as an initial state with a bound state approximation.
For other resonances of $^7$He, their decay widths are more than 1 MeV, as shown in Table~\ref{ene7}, 
so that the bound state approximation is not considered to be valid. 
When the removal strength of these broad $^7$He resonances are calculated, 
all the continuum components of $^7$He, such as $^6$He+$n$ $^5$He+$2n$, $^4$He+$3n$, should be included simultaneously in the strength.
However, this treatment is difficult to carry out at the present stage, and further theoretical development is desired.

\section*{ACKNOWLEDGMENTS}
This work was supported by a Grant-in-Aid for Young Scientists from the Japan Society for the Promotion of Science (JSPS, No. 21740194).
Numerical calculations were performed on the computer system at RCNP, Osaka University.

\end{document}